\documentclass[referee]{raa}

\usepackage[a4paper,margin=2.5cm]{geometry}
\usepackage{graphicx,times}
\usepackage{natbib}
\usepackage{amssymb,amsmath}
\bibpunct{(}{)}{;}{a}{}{,}

\usepackage[pagebackref=true]{hyperref}

\usepackage{bm}

\begin{document}

  \title{Redshift Dependence of the Low-energy Spectral Index
  	of Gamma-Ray Bursts Revisited}

   \volnopage{Vol.0 (20xx) No.0, 000--000}      
   \setcounter{page}{1}          

   \author{Xiao-Li Zhang 
      \inst{1}
   \and Yong-Feng Huang 
      \inst{2,3,4}
   \and Ze-Cheng Zou 
      \inst{2}
   }

   \institute{Department of Physics, Nanjing University, Nanjing 210023, People's Republic of China\\
        \and
             School of Astronomy and Space Science, Nanjing
             University, Nanjing 210023, People's Republic of China; {\it hyf@nju.edu.cn}\\
        \and
             Key Laboratory of Modern Astronomy and Astrophysics
             (Nanjing University), Ministry of Education, People's Republic of
             China\\
        \and
             Xinjiang Astronomical Observatory, Chinese Academy of Sciences, Urumqi 830011, People's Republic of China\\
\vs\no
   {\small Received 20xx month day; accepted 20xx month day}}

\abstract{ A negative correlation was found to exist between the low-energy spectral index and the redshift of gamma-ray bursts (GRBs) by \citet{Amati2002}. It was later confirmed by \citet{Geng2013} and \citet{7-DavidGruber2014}, but the correlation was also found to be quite dispersive when the sample size was significantly expanded. In this study, we have established two even larger samples of gamma-ray bursts to further examine the correlation. One of our sample is consisted of 316 GRBs detected by the Swift satellite, and the other one is consisted of 80 GRBs detected by the Fermi satellite. It is found that there is no correlation between the two parameters for the Swift sample, but there does exist a weak negative correlation for the Fermi sample. The correlation becomes even more significant when the spectral index at the peak flux is considered. It is argued that the absence of the correlation in the Swift sample may be due to the fact that Swift has a very narrow energy response so that it could not measure the low-energy spectral index accurately enough. Further studies based on even larger GRB samples are solicited.
}


\maketitle

%
%
\section{INTRODUCTION}           
\label{INTRODUCTION}

Gamma-ray bursts (GRBs) are the most violent stellar explosions. They were initially discovered in 1967 \citep{1-1-klebesadel1973} by the Vela Satellites. Up to now, they have been intensively studied for more than forty years since more and more GRBs are discovered \citep{Qin_2021, Liu_2022, Yuan_2022}. Generally, GRBs are believed to be triggered by the death of massive stars or by the merger of binary compact stars. The empirical correlation between different parameters of GRBs is one of the present interesting objects. For instance, the correlation between $E_p$(the peak energy) and $E_{iso}$(the equivalent isotropic energy), i.e., the so-called Amati relation is investigated by \cite{Amati2002,Amati2009},
\cite{Virgili2012}, \cite{Geng2013}, \cite{9-WalidJ.Azzam2013} and \cite{Demianski2017}. The correlation between $E_p$(the peak energy) and $L_p$(the peak luminosity), i.e., the so-called Yonetoku relation is investigated by \cite{Yonetoku2004}, \cite{Ghirlanda2005} and \cite{15-Z.B.Zhang2012}. \cite{YONETOKU2010} and \cite{Tsutsui2013} have discussed both the two relations. Furthermore, \cite{1-FeifeiWang2020} has even discussed other more empirical correlations beyond above relations.

The redshift ($z$) is an important parameter of GRBs. Many people have
studied the relation between $z$ and other parameters. For example,
\cite{2-FuWenZhang2014} investigated the correlations between $z$ and
$L_{iso}$, $E_{iso}$, $E_{p,rest}$, $E_{p,obs} $ ($E_{p,obs} = E_{p,rest}/(1 + z)$)
(also see \citet{2-1-sakamoto2011,2-2-ukwatta2011}).  Especially, they
found a positive correlation between $z$ and $L_{iso}$ for the Swift
GRBs.  \cite{16-D.M.Wei2003} and \cite{8-H.Zitouni2014} studied the
$z - E_{iso}$ relation and the $z-L_{iso}$ relation. Moreover, the
relation between $z$ and $L_p$ has also been investigated by many
researchers \citep{4-1-NicoleM.Lloyd-Ronning2002,3-1-salvaterra2012,
	4-2-GOLDSTEIN2012,3-Z.B.Zhang2014,4-H.Zitouni2018}.

The low-energy spectral index, $\alpha$, is also an important parameter
that characterizes the prompt emission of GRBs. \cite{6-Jin-JunGeng2018}
have tried to derive the value of $\alpha$ through numerical simulations
by considering the synchrotron emission mechanism. Especially, the
correlation between $\alpha$ and $E_p$ has been studied by many
researchers \citep{7-DavidGruber2014,11-LiangLi2019,5-Ming-YaDuan2020,12-LiangLi2022}.
\cite{13-Chen-HanTang2019} also investigated the correlation
between $\alpha$ and $E_{iso}$.

Interestingly, \cite{Amati2002} found that there seems to be a negative
correlation between $z$ and $\alpha$, which
reads $\log\alpha = (-0.78\pm0.13) \log(1+z)+(0.39\pm0.04)$.
Here the symbol $log$ means the logarithm base is 10. Their
study is conducted on a sample that only includes 9 GRBs.
Later, \cite{Geng2013} and \cite{7-DavidGruber2014} confirmed the existence of
such a correlation, which was updated
as $\log\alpha = (-0.42 \pm 0.07) \log(1 + z) + (0.11 \pm 0.02)$.
Their sample is consisted of 65 GRBs, which is much larger than that
of \cite{Amati2002}. However, \cite{Geng2013} also noted that the
correlation is quite dispersive. It is thus necessary to further
examine the existence of such a $z - \alpha$ correlation. In this
study, we will further expand the GRB sample to include 316 GRBs
detected by Swift and 116 GRBs detected by Fermi. We then use this
significantly expanded sample to check the correlation between $z$ and
$\alpha$.

The structure of this article is organized as follows. A detailed
description on the selection of the data sample is provided in
Section \ref{SAMPLE}. The correlation between the low-energy spectral
index and the redshift, and some other relations between various parameter
pairs, are explored in Section \ref{CORRELATION}. Finally, our conclusions
and discussion are presented in Section \ref{DISCUSSION AND CONCLUSIONS}.


\section{SAMPLE}
\label{SAMPLE}

GRBs detected by Swift and Fermi are used in this study. Two conditions
are applied in selecting the appropriate GRBs.
First, the redshift of the burst should be available.
Second, the spectrum should be well defined.
The time-averaged spectra of GRBs are usually fitted with three kinds of functions:
a single power-law function, a cutoff power-law function, and the so called Band
Function \citep{Band1993}. Most of the GRBs spectra can be well fitted by the Band
function, which is expressed as
\begin{eqnarray}\label{func1}
	F(E) = A \begin{cases}
		(\frac{E}{100 keV}) ^{-\alpha} \exp\left[  -\dfrac{(2-\alpha ) E}{E_p}\right] , & E < \frac{-(\alpha - \beta) E_p}{2-\alpha}; \\
		\left(\frac{E}{100 keV}\right) ^{-\beta} \exp(\alpha-\beta) \left[\frac{-(\alpha-\beta) E_p}{100 keV (2-\alpha)} \right] ^{-(\alpha-\beta)}, & E \geqslant \frac{-(\alpha - \beta) E_p}{2-\alpha},
	\end{cases}
\end{eqnarray}
where $E$ is the photon energy.
There are four parameters in this equation: the scaling factor ($A$), the low-
and high- energy spectral indices ($\alpha$ and $\beta$, respectively), and
the peak photon energy ($E_p$). In our notation, $\alpha$ and $\beta$ are
positive by using their absolute values. We have collected all the spectrum parameters
of those Swift and Fermi GRBs with the redshift measured.

Fermi has a very wide energy band, i.e. 8 keV -- 35,000 keV.
So, Fermi GRBs are usually best fitted by the Band function,
which is represented by the lower-energy spectral index ($\alpha$)
and the high-energy spectral index ($\beta$). On the other hand,
the Swift/BAT detector has a very narrow energy response
of 15 keV -- 150 keV so that Swift GRBs are generally best
fitted by a single power-law function or by a cutoff
power-law function. In these cases, the derived power-law index could
be regarded as a useful representation of the low-energy spectral
index ($\alpha$), since it is measured in the soft $\gamma$-ray range.
Note that for the Swift GRBs, the $\beta$ parameter is completely
unavailable.

For the Swift GRBs, the relevant data are acquired by inquiring the NASA Swift website\footnote{\href{https://swift.gsfc.nasa.gov/archive/}{https://swift.gsfc.nasa.gov/archive/}\label{web1}}.
As a result, a total number of 316 GRBs are included in our Swift sample,
all of which have the necessary redshift and spectrum data.
The time span of these GRBs ranges from Jan 26, 2005 to Jan 16, 2023.

For the Fermi GRBs, the data are collected mainly by consulting NASA's HEASARC
(the High Energy Astrophysics Science Archive Research Center)
database\footnote{\href{https://heasarc.gsfc.nasa.gov/db-perl/W3Browse/}{https://heasarc.gsfc.nasa.gov/db-perl/W3Browse/}\label{web2}}.
As a useful supplement, 24 bursts were taken from Jochen Greiner's
online GRB catalog \footnote{\href{https://www.mpe.mpg.de/~jcg/grbgen.html}{https://www.mpe.mpg.de/~jcg/grbgen.html}\label{web3}}.
Finally, our Fermi sample is consisted of 80 GRBs.
The time span of these Fermi GRBs ranges from Sep 05, 2008 to July 20, 2018.

\section{CORRELATIONS}
\label{CORRELATION}

\begin{figure}[htbp]
	\includegraphics[scale=0.45]{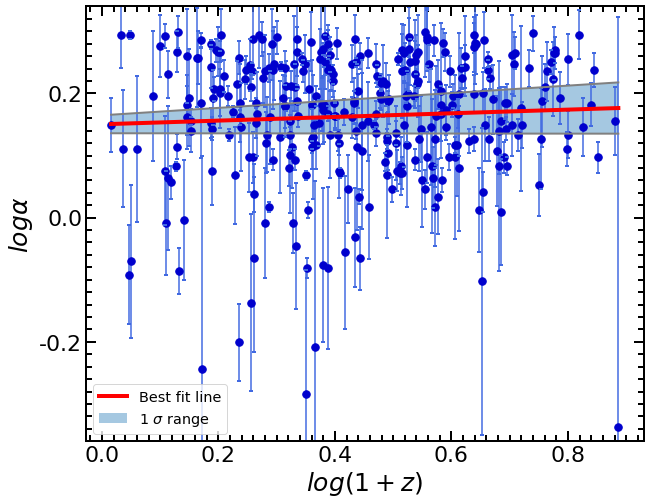}
	\centering
	\caption{Swift sample on the $\log\alpha-\log(1+z)$ plane.
		Here $\alpha$ is the low-energy spectral index and $z$
		is the redshift.
        The solid line is the best fit result
        and the shaded region shows the corresponding $1 \sigma$ range.}
	\label{fig1}
\end{figure}

In this section, the Swift and Fermi GRB samples are employed to
explore the empirical correlations concerning $\alpha$ and $z$.
Figure~\ref{fig1} shows the Swift sample on the $\log\alpha-\log(1+z)$ plane.
Generally, we see that the data points are quite scattered and there is no
obvious correlation between the two parameters. The solid line in Figure~\ref{fig1}
is the best-fit result of the data points, which corresponds
to $\log\alpha= (0.029\pm0.030) \log(1+z) + (0.15\pm0.014)$. But note that it has a very small
correlation coefficient of $r= 0.055$, which indicates that essentially no
correlation exists.
Here, the error ranges of the best-fit parameters are given in $1 \sigma$ ranges throughout this paper. 

\begin{figure}[htbp]
	\centering
	\begin{minipage}{0.49\textwidth}
		\centering
		\includegraphics[scale=0.31]{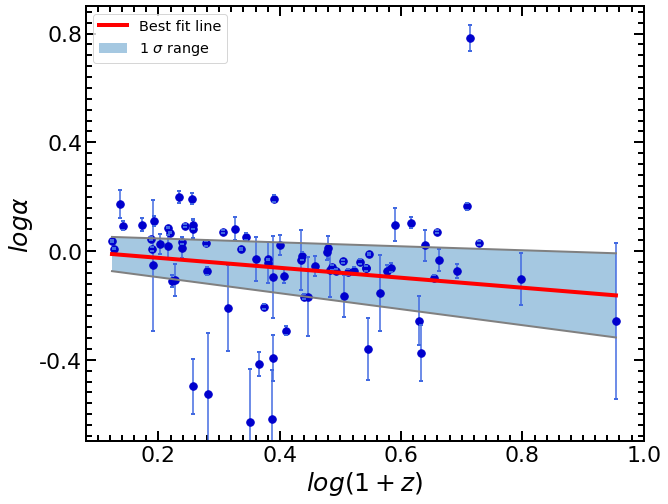}
	\end{minipage}
	\begin{minipage}{0.49\textwidth}
		\centering
		\includegraphics[scale=0.31]{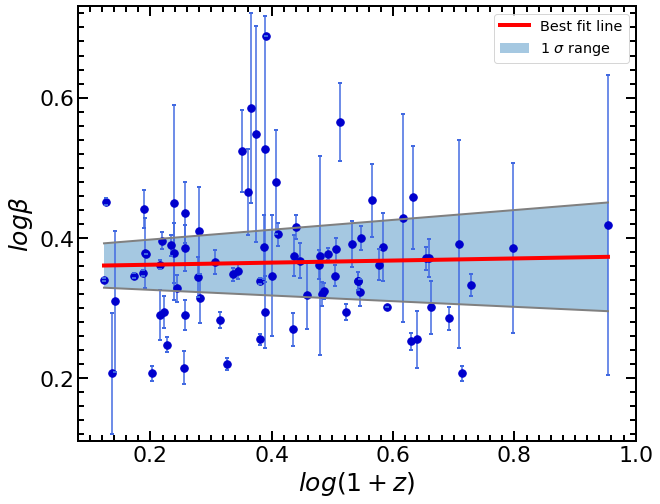}
	\end{minipage}
	\caption{The spectral index plotted against
		    redshift for the
			Fermi GRB sample. The left panel shows the low-energy spectral index ($\alpha$)
			versus redshift ($z$). The right panel shows the high-energy spectral
			index ($\beta$) versus redshift ($z$). The solid lines correspond to
			the best fit results to the data points and the shaded regions show the corresponding $1 \sigma$ ranges. Note that in
			the left panel, the isolated data point (GRB 120712A) on the top is not included in the linear fit since it is an obvious outlier.}
	\label{fig2}
\end{figure}

In Figure \ref{fig2}, we plot the Fermi GRB sample on the $\log\alpha-\log(1+z)$ plane
in the left panel. The observational data points are also fitted with a linear function. Note that
GRB 120712A seems to be an obvious outlier on the plane, whose low-energy spectral index is
$\alpha = 6.1 \pm 0.67$. We exclude this event in the fitting procedure.
Interestingly, we could see that there is a negative correlation
between $\alpha$ and $z$. The best-fit result
is $\log\alpha = (-0.18 \pm0.11) \log(1+z) +(0.012\pm0.049)$,
with a correlation coefficient of ${r = -0.19}$. But note that this expression, especially
the slope of the line (i.e. ${-0.18}$), is quite different from those of previous studies.
Also, the correlation here is generally very weak.
As a comparison, the right panel of Figure \ref{fig2} shows the Fermi GRB sample on the $\log\beta-\log(1+z)$ plane. We see that there is no correlation between the two parameters, which is consistent with previous results of \citet{Geng2013}.
The best-fit line corresponds to $\log\beta = (0.015\pm0.055) \log(1+z)+ (0.36\pm0.025)$,
with a very small correlation coefficient of ${r = 0.031}$.
Comparing the two panels of Figure \ref{fig2}, it is quite clear that the low energy
spectral index ($\alpha$) is still obviously different from the high energy index
($\beta$) when they are plotted against the redshift.

\begin{figure}[htbp]
	\centering
	\begin{minipage}{0.49\textwidth}
		\centering
		\includegraphics[scale=0.31]{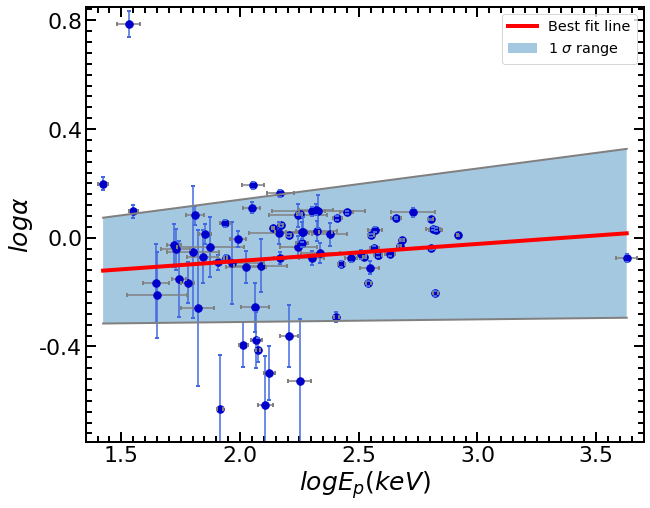}
	\end{minipage}
	\begin{minipage}{0.49\textwidth}
		\centering
		\includegraphics[scale=0.31]{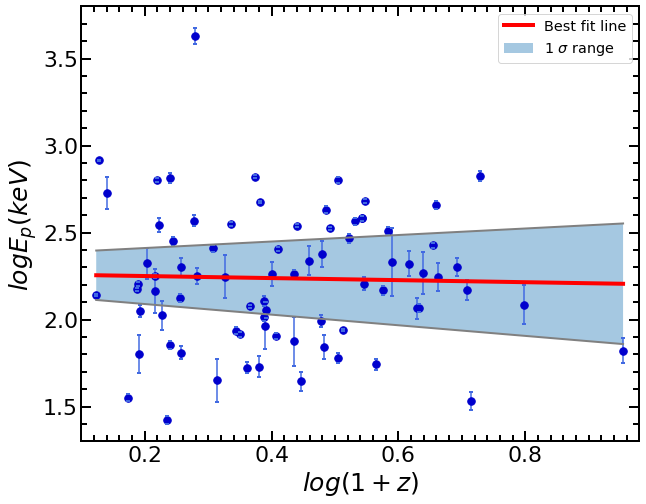}
	\end{minipage}
	\caption{$\alpha$ versus $E_p$ (left panel) and
		    $E_p$ versus $z$ (right panel) for the Fermi GRBs.
            The solid lines correspond to
			the best fit results to the data points and the shaded regions show the corresponding $1 \sigma$ ranges. Note that in
			the left panel, the isolated data point (GRB 120712A) on the top is not included in the linear fit since it is an obvious outlier.}
	\label{fig3}
\end{figure}

The Fermi GRBs have well measured $E_p$ data. So, in Figure \ref{fig3}, we
plot them on the $\alpha - E_p$ plane and the $E_p - z$ plane.
We see that there is no obvious correlation either between $\alpha$ and
$E_p$, or between $E_p$ and $z$. The results are also somewhat different
from those of previous studies. For example, \citet{Geng2013}
argued that $E_p$ and $z$ are positively correlated.
Note that GRB 120712A is again an obvious outlier in this
figure and is not included in the fitting procedure.

Since the peak time of the flux is an important stage of a GRB, we have also
investigated the features of the peak flux parameters. For this purpose,
we could only adopt the Fermi GRB sample, since peak flux parameters
are not available for many Swift GRBs. Here, the low-energy spectral index
at the peak flux of the burst is denoted as $\alpha_{\rm peak}$. Correspondingly,
the high-energy spectral index and peak photon energy at the peak flux are
denoted as $\beta_{\rm peak}$ and $E_{\rm peak}$, respectively.

\begin{figure}[htbp]
	\centering
	\begin{minipage}{0.49\textwidth}
		\centering
		\includegraphics[scale=0.31]{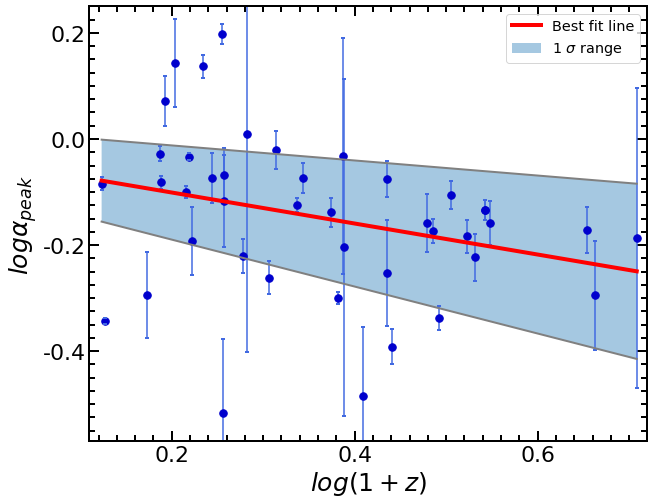}
	\end{minipage}
	\begin{minipage}{0.49\textwidth}
		\centering
		\includegraphics[scale=0.31]{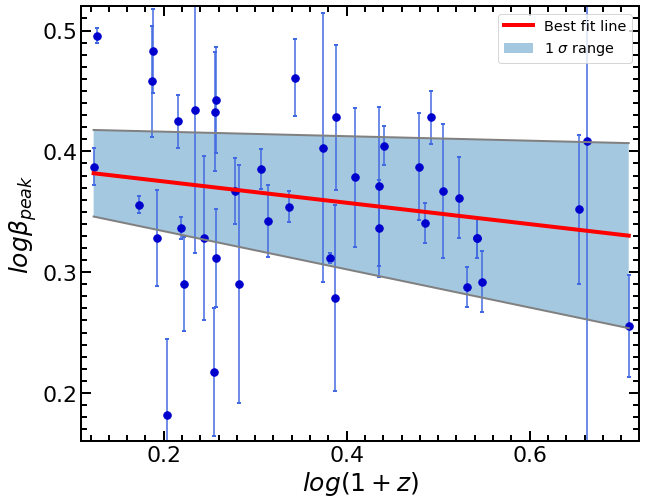}
	\end{minipage}
	\caption{The peak flux spectral indices plotted against the redshift for the
		Fermi GRB sample. The left panel shows the low-energy spectral
		index at the peak flux moment ($\alpha_{\rm peak}$)
		versus the redshift. The right panel shows the corresponding high-energy spectral
		index ($\beta_{\rm peak}$) versus the redshift. The solid lines correspond to
		the best fit results to the data points
        and the shaded regions show the corresponding $1 \sigma$ ranges.}
	\label{fig4}
\end{figure}

In the left panel of Figure \ref{fig4}, the Fermi GRB sample are plotted
on the $\log\alpha_{\rm peak}-\log(1+z)$ plane. Here, we could see that there is an
obvious negative correlation between $\alpha_{\rm peak}$ and $z$, although it is
still quite scattered. The best-fit result
is $\log\alpha_{\rm peak} = (-0.29\pm0.15) \log(1+z)+ (-0.042\pm0.059)$,
with a correlation coefficient of $r = -0.29$.
Comparing with the slope of $-0.42$ previously derived for
the $\log\alpha - \log(1+z)$ correlation \citep{Geng2013}, here the slope
of the $\log\alpha_{\rm peak} - \log(1+z)$ relation (i.e. $-0.29$) is significantly
flatter. The right panel of Figure \ref{fig4} shows the Fermi GRB sample on
the $\log\beta_{\rm peak}-\log(1+z)$ plane. There is also a
weak correlation between the two parameters, which can be best
fitted as $\log\beta_{\rm peak} = (-0.088\pm0.070) \log(1+z) + (0.39\pm0.027)$, with a
correlation coefficient of $r = -0.20$. Note that the data points
in the right panel are much dispersive as compared with those in
the left panel.

\begin{figure}[htbp]
	\includegraphics[scale=0.45]{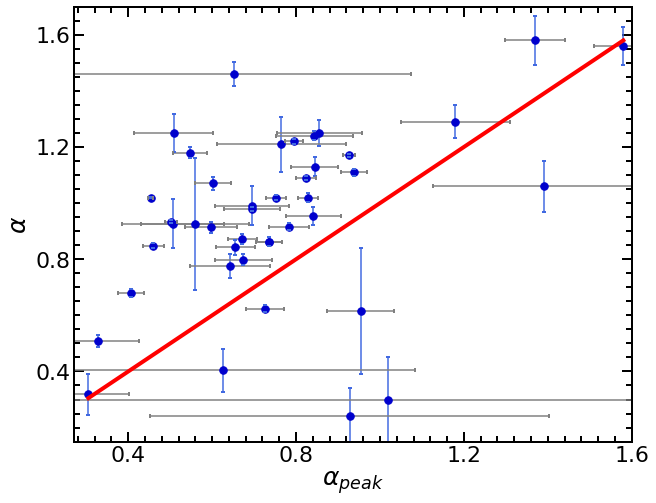}
	\centering
	\caption{$\alpha_{\rm peak}$ plotted versus $\alpha$ for the Fermi GRB sample. The solid line corresponds to the case of $\alpha = \alpha_{\rm peak}$.}
	\label{fig5}
\end{figure}

Figures \ref{fig2} and \ref{fig4} show that there is a weak correlation between
the low-energy spectral index and the redshift. Especially, the $\alpha - z$
relation is much less significant than the $\alpha_{\rm peak} - z$ relation.
A natural speculation is that $\alpha$ and $\alpha_{\rm peak}$ should be
positively connected, then the above two relations should be largely similar.
The reason that leads to such a difference thus deserves to be examined.
Figure \ref{fig5} plots $\alpha$ against $\alpha_{\rm peak}$ for the
Fermi sample. The solid line shows the case when $\alpha$
equals $\alpha_{\rm peak}$. We see that $\alpha$ and $\alpha_{\rm peak}$
are not strictly connect, which could explain their different dependence
on the redshift. It reflects the fact that the $\gamma$-ray spectrum is
highly variable during a GRB.

\begin{figure}[htbp]
	\centering
	\begin{minipage}{0.49\textwidth}
		\centering
		\includegraphics[scale=0.31]{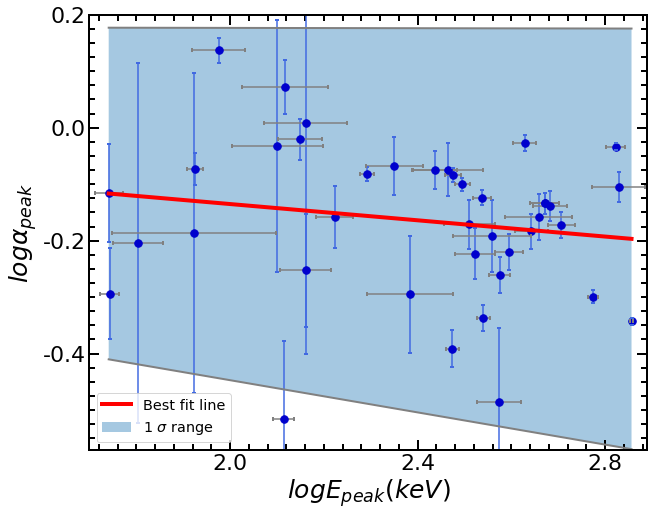}
	\end{minipage}
	\begin{minipage}{0.49\textwidth}
		\centering
		\includegraphics[scale=0.31]{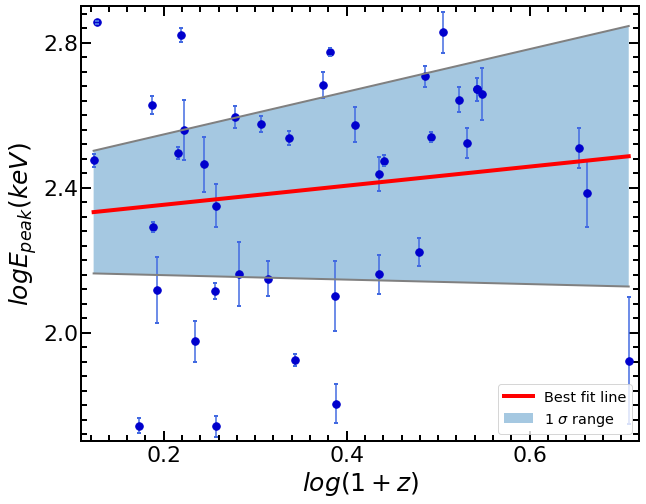}
	\end{minipage}
	\caption{$\alpha_{peak}$ plotted versus $E_{peak}$ (left panel), and $E_{peak}$
		plotted versus $z$ (right panel) for the Fermi GRB sample. The solid lines
		correspond to the best fit results
        and the shaded regions show the corresponding $1 \sigma$ ranges.} 
	\label{fig6}
\end{figure}

Figure \ref{fig6} illustrates the relations between $\alpha_{\rm peak}$,
$E_{\rm peak}$ and $z$. There is a weak correlation between $E_{\rm peak}$
and $z$, which reads $\log E_{\rm peak} = (0.26\pm0.33) \log(1+z) + (2.30\pm0.13)$, with
the correlation coefficient being $r =0.13$. Similarly, there is
also a weak correlation between $\alpha_{\rm peak}$ and $E_{\rm peak}$,
which corresponds to $\log \alpha_{peak} = (-0.072\pm0.071) \log E_{peak} + (0.0088\pm0.17)$,
with the correlation coefficient being $r = -0.16$.

\begin{figure}[htbp]
	\centering
	\includegraphics[scale=0.45]{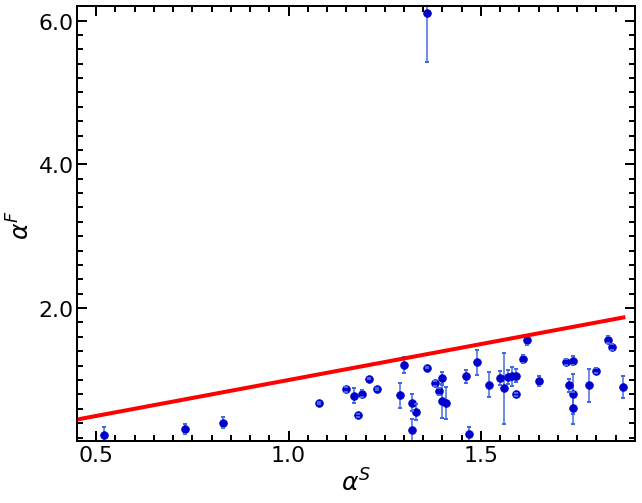}
	\caption{The low-energy spectral index measured by Fermi ($\alpha^F$),
			plotted versus the spectral index measured by Swift ($\alpha^S$).
			Note that only overlapping GRBs between the Fermi sample and
			the Swift sample are shown here.
			The solid line corresponds to the case of $\alpha^F = \alpha^S$.}
	\label{fig7}
\end{figure}

\begin{figure}[htbp]
	\includegraphics[scale=0.45]{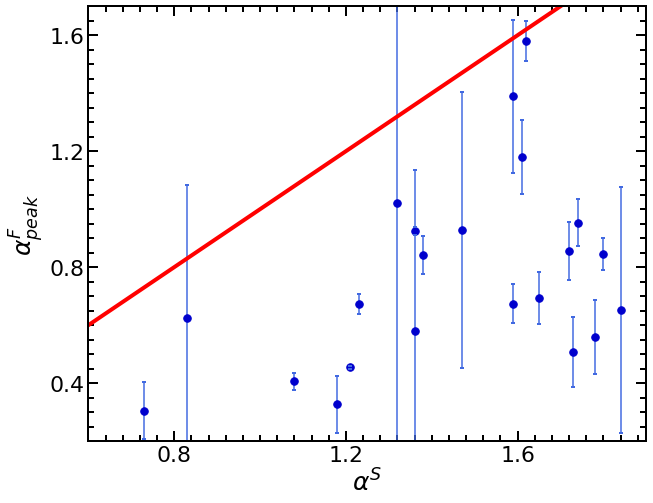}
	\centering
	\caption{The low-energy spectral index at the peak flux measured by
		Fermi ($\alpha_{\rm peak}^F$), plotted versus the spectral index
		measured by Swift ($\alpha^S$). Note that only overlapping GRBs between
		the Fermi sample and the Swift sample are shown here.
		The solid line corresponds to the case of $\alpha_{\rm peak}^F = \alpha^S$.}
	\label{fig8}
\end{figure}

Swift/BAT has a relatively narrow passband, i.e. 15 keV -- 150 keV. Therefore, the
measured spectral index should correspond to the low-energy spectral index of the
Band function. Some GRBs are simultaneously detected by Swift and Fermi satellites.
It is then interesting to know whether the spectral index measured by Swift is
consistent with that measured by Fermi. In Figure \ref{fig7}, we have screened out
all the overlapping GRBs between the Swift sample and Fermi sample, and compared
their spectral indices. Here, $\alpha^S$ is the spectral index measured by
Swift, while $\alpha^F$ is the low-energy spectral index measured by Fermi.
We see that $\alpha^S$ and $\alpha^F$ are not equal for each event.
The former is systematically larger than the latter. Also, the data points
are quite scattered. For these overlapping GRBs, we have also compared their
$\alpha^S$ with their low-energy spectral index at the peak flux as measured
by Fermi ($\alpha_{\rm peak}^F$). The results are shown in Figure \ref{fig8}.
Similarly, we see that $\alpha^S$ again is generally larger than $\alpha_{\rm peak}^F$.
Figures \ref{fig7} and \ref{fig8} clearly shows that different detectors could
generate very different results for the spectrum of even the same GRB, which
indicates that acquiring the spectra of GRBs is still a very difficult task.
The different $\alpha-z$ correlations of different GRB samples could
thus be caused by the systematic distortion in measuring the spectral indices.

\section{DISCUSSION AND CONCLUSIONS}
\label{DISCUSSION AND CONCLUSIONS}

In this study, we use the Swift GRBs and Fermi GRBs to explore the possible
correlation between the low-energy spectral index and the redshift. For the
Swift GRB sample, it is found that there is no any correlation between $\alpha$
and $z$ (see Figure \ref{fig1}). On the other hand, there is a weak correlation
between $\alpha$ and $z$ for the Fermi GRB sample (see Figure \ref{fig2}). The
correlation is even more obvious when the peak flux spectrum is considered, i.e.
when $\alpha_{\rm peak}$ is plotted versus $z$ (Figure \ref{fig4}). The different
features of the two samples may be caused by the different energy response of
the two detectors. The energy band of Fermi is very wide (8 keV -- 35,000 keV),
which ensures that it can present a much better description of the
spectra of the detected GRBs. However, the Swift/BAT has a very narrow energy
response (15 keV -- 150 keV). As a result, the spectra of Swift GRBs are usually
best fitted by a single power law function or by a cutoff power-law function. In
these cases, the power-law spectral index may significantly deviate from the true
low-energy spectral index. This conjecture was confirmed when the spectra of the
overlapping GRBs of the two samples were scrutinized. It is found that when a GRB
is simultaneously detected by both Fermi and Swift, then the spectral index reported
by Swift is usually quite different from the low-energy spectral index ($\alpha$)
reported by Fermi (Figures \ref{fig7} \& \ref{fig8}). It reflects the difficulty
in spectral observations of GRBs.

\citet{Amati2002} argued that the correlation between $\alpha$ and $z$
is $\log\alpha = (-0.78\pm0.13) \log(1+z)+(0.39\pm0.04)$. Note that their
sample size is very small. Later, \citet{Geng2013} updated the correlation
as $\log\alpha = (-0.42 \pm 0.07) \log(1 + z) + (0.11 \pm 0.02)$ using a
significantly expanded sample. We see that the slope between $\log\alpha$
and $\log(1+z)$ becomes much flatter, and the correlation is also very
dispersive. According to our current study, the correlation is now
$\log\alpha = (-0.18\pm0.11) \log(1 + z) + (0.012\pm0.049)$ for the Fermi sample, and it
is $\log \alpha_{\rm peak} = (-0.29\pm0.15) \log(1+z) +(-0.042\pm0.059)$ for the peak
flux spectrum. The correlation becomes even much flatter and more
dispersive. Anyway, based on this study, we can conclude that a weak
correlation does exist between $\alpha$ and $z$. Also, further study
on this issue still deserves being conducted in the future. For this
purpose, detectors with a relatively wide energy response are
necessary to accurately measure the exact spectra of GRBs. We note
that the Einstein Probe, a wide field (3600 square degree) X-ray
satellite that is scheduled to be launched in late 2023 \citep{WeiminYuan2018,WeiminYuan2022}, might
be very helpful in this aspect. Einstein Probe is very sensitive
in X-rays and may markedly help increase the sample size of
well-localized GRBs.

\begin{acknowledgements}
	This study was supported by the National Natural Science Foundation
	of China (Grant Nos. 12233002, 12041306, 12147103, U1938201), by National SKA
	Program of China No. 2020SKA0120300, by the National Key R\&D
	Program of China (2021YFA0718500),
	and by the Youth Innovations and Talents Project of Shandong Provincial
	Colleges and Universities (Grant No. 201909118).
	
\end{acknowledgements}

\bibliography{ref}{}
\bibliographystyle{raa}

\label{lastpage}

\end{document}